\documentclass{PoS}
\usepackage{amsmath}
\usepackage{amssymb}
\usepackage{bbm}
\usepackage{epsfig}
\usepackage{slashed}

\def\ga{\mathrel{\raise.3ex\hbox{$>$\kern-.75em\lower1ex\hbox{$\sim$}}}}
\def\la{\mathrel{\raise.3ex\hbox{$<$\kern-.75em\lower1ex\hbox{$\sim$}}}}

\def\lsim{\mathrel{\rlap{\lower4pt\hbox{\hskip1pt$\sim$}}
    \raise1pt\hbox{$<$}}}                
\def\gsim{\mathrel{\rlap{\lower4pt\hbox{\hskip1pt$\sim$}}
    \raise1pt\hbox{$>$}}}                

\title{BSM Higgs searches in the gluon fusion process $pp \to h +jet \to \tau^+ \tau^-  + jet$ at the LHC}

\ShortTitle{BSM Higgs searches in $pp \to h +jet \to \tau^+ \tau^-  + jet$}

\author{Alexander Belyaev\\
        NExT Institute and School of Physics and Astronomy,
        University of Southampton, Highfield, Southampton SO17 1BJ, UK.\\
        Particle Physics Department, Rutherford Appleton Laboratory, Chilton, Didcot, Oxon OX11 0QX, UK.\\
        E-mail: \email{a.belyaev@soton.ac.uk}}

\author{Renato Guedes\\
        NExT Institute and School of Physics and Astronomy,
        University of Southampton, Highfield, Southampton SO17 1BJ, UK.\\
        E-mail: \email{r.b.guedes@soton.ac.uk}}

\author{Stefano Moretti\\
        NExT Institute and School of Physics and Astronomy,
        University of Southampton, Highfield, Southampton SO17 1BJ, UK.\\
        Particle Physics Department, Rutherford Appleton Laboratory, Chilton, Didcot, Oxon OX11 0QX, UK.\\
        E-mail: \email{stefano@phys.soton.ac.uk}}

\author{\speaker{Rui Santos\thanks{I acknowledge the warm hospitality of the University of Toyama, Japan, from where this talk was delivered (courtesy of Eyjafjallajökull). RG is supported by FCT, grant SFRH/BPD/47348/2008.
RS is supported by the FP7 via a Marie Curie IEF,  PIEF-GA-2008-221707.}}\\
        NExT Institute and School of Physics and Astronomy,
        University of Southampton, Highfield, Southampton SO17 1BJ, UK.\\
        Particle Physics Department, Rutherford Appleton Laboratory, Chilton, Didcot, Oxon OX11 0QX, UK.\\
        E-mail: \email{rsantos@cii.fc.ul.pt}}

\abstract{The mass of the Standard Model (SM) Higgs boson was constrained by the Large Electron-Positron (LEP) collider to be above 114.4 $GeV$. Simple extensions of the scalar sector like adding singlets and/or doublets allow the lightest Higgs to be much lighter via a reduction in its coupling to gauge bosons. Such a light Higgs could have evaded the LEP searches but could be detected at the Large Hadron Collider (LHC) in the process $pp \to h j \to \tau^+ \tau^- j$, where $j$ is a resolved jet. We conclude that in some Beyond the Standard Model (BSM) extensions of the scalar sector, a very light Higgs state could be detected at the 14 TeV LHC with early data.}

\FullConference{XVIII International Workshop on Deep-Inelastic Scattering and Related Subjects, DIS 2010\\
		April 19-23, 2010\\
		Firenze, Italy}

\begin{document}

\section{Introduction}
\label{sec:intr}

The Higgs mechanism gives rise to clear and detectable signatures that can be probed at hadron colliders. The most relevant searches for a SM Higgs boson by the LEP experiments~\cite{Schael:2006cr} were based on the process $e^+ e^- \to Z h$. A limit of $m_h > 114.4$ $GeV$ was obtained combining all LEP analyses based on such searches. As the coupling of a Higgs boson to gauge bosons is fixed in the SM, a lighter Higgs can only exist in a model where its couplings to gauge bosons are reduced relative to the SM. Moreover, any complementary process, such as $e^+e^-\to Ah$, where $A$ is a pseudo-scalar Higgs boson and $h$ is the lightest Higgs, has to be kinematically forbidden. Such a scenario, with a very light Higgs, can easily arise by adding Higgs scalar singlets and/or doublets to the Higgs sector of the SM. This way, a reduction of the couplings to gauge bosons can occur via the mixing of the different Higgs states which in turn would lead to negligible Higgs production cross sections involving such couplings. Hence, the detection of a very light Higgs boson necessarily involves its Yukawa couplings. The process $pp \to h j \to \tau^+ \tau^- j$~\cite{Ellis:1987xu} via gluon fusion, where $j$ represents a resolved jet, proved to be the most important in this limiting scenario. A detailed parton level study was performed for the LHC with a center-of-mass energy of $\sqrt{s} =14$ TeV, showing that such a light Higgs could be detected in several extended models and that for particular scenarios an early detection is also possible. We believe that at the very least, an effort should be made to find or definitely exclude such a light particle and the LHC has the means to do so.

\section{Extensions of the Higgs sector}
\label{sec:ext}

Which extensions of the SM can accommodate a very light Higgs boson? For simplicity we assume that CP is conserved in the Higgs sector and natural flavour conservation is  also assumed. The addition of a neutral scalar singlet would at most redefine the Higgs coupling to gauge bosons as $\sin \chi \, g_{VVh}^{SM}$ because this field does not couple to gauge bosons nor to fermions - $\chi$ is the mixing angle between the CP-even component of the singlet and the CP-even Higgs field from the doublet. However, the coupling to fermions will also be redefined as $\sin \chi \, g_{ffh}^{SM}$. Consequently, a small $\sin \chi$ will make all production processes negligible and detection of this particular very light Higgs does not seem promising at the LHC.

The next step is to add one doublet to the SM to obtain what is known as a two-Higgs doublet model (2HDM). The Higgs couplings to gauge bosons are universal and the lightest Higgs state couples as $\sin (\beta - \alpha) \, g_{SM}$, where $\beta$ is the mixing angle in the CP-odd and charged sectors and $\alpha$ is the mixing angle in the CP-even Higgs sector. The Yukawa Lagrangian can be built in four independent ways~\cite{catalogue} if Flavour Changing Neutral Currents (FCNCs) are to be avoided. The first one is a SM-like scenario where only one doublet, say $\phi_2$, gives mass to all fermions usually referred to as type I model. The second class of models is the one where both doublets participate in the mass generation process. One can build the following models: type II is the model where $\phi_2$ couples to up-type quarks and $\phi_1$ couples to down-type quarks and leptons; in a type III model $\phi_2$ couples to up-type quarks and to leptons and $\phi_1$ couples to down-type quarks; a type IV model is instead built such that $\phi_2$ couples to all quarks and $\phi_1$ couples to leptons.

A class of simple extensions of 2HDM recently discussed in~\cite{Barger:2009me} are the ones obtained by adding an arbitrary number, $n$, (we will call these models 2HDM+nD) of doublets that \textit{do not couple} to the fermions. Taking Model II as an example and following~\cite{Barger:2009me} we discuss the case where just one more doublet is added; the mixing with the extra doublet is parameterised in terms of an angle $\theta$, where $h= \cos \theta h' + \sin \theta h_0$ and $h'$ is the usual 2HDM lightest CP-even Higgs boson and $h_0$ is the CP-even state of the new doublet. Defining $\tan \beta = v_2/v_1$, $\cos \Omega =\sqrt{(v_1^2 + v_2^2)}/v$ and $\sin \Omega = v_0 /v$, with $0 \leq \Omega < \pi/2$, the couplings to gauge bosons are
\begin{equation}
g_{VVh} = (\cos \Omega \cos \theta \sin (\beta - \alpha) + \sin \Omega \sin \theta) \, g_{VVh}^{SM}
\end{equation}
while the couplings to fermions can be written as
\begin{equation}
g_{\bar{u} u h} = \frac{\cos \theta}{\cos \Omega} \frac{\cos \alpha}{\sin \beta} \, g_{ffh}^{SM}, \qquad \qquad g_{\bar{d} d h} = g_{\bar{l} l h} = - \frac{\cos \theta}{\cos \Omega} \frac{\sin \alpha}{\cos \beta} \, g_{ffh}^{SM}.
\end{equation}
In the case of the so-called democratic three-Higgs doublet model 3HDM(D)~\cite{Grossman:1994jb} where up-type quarks, down-type quarks and charged leptons all get their mass from a different doublet we can still write the coupling to gauge bosons as in eq (3.1), where now $\tan \beta = v_u/v_d$, $\cos \Omega =\sqrt{(v_u^2 + v_d^2 )}/v$ and $\sin \Omega = v_l /v$; $v_u$, $v_d$ and $v_l$ are the VEVs of the doublets that couple to the up-quarks, down-quarks and charged leptons, respectively. Couplings to fermions have the same expression except for the coupling to leptons which is now given by $\sin \theta/\sin \Omega \enskip g_{ffh}^{SM}$.

\section{Results and discussion}

\begin{table}[ht]
\begin{center}
\begin{tabular}{c c c c c c c c c c} \hline \hline
Mass ($GeV$) & 95 \% CL  $L~(fb^{-1})$ && 3$\sigma$ $L~ (fb^{-1})$&&  5$\sigma$ $L~ (fb^{-1})$ \\
\hline
20        &  0.38     &&    0.86  &&   2.38   \\ \hline
30        &  0.45     &&    1.02  &&   2.84   \\ \hline
40        &  0.53     &&    1.18  &&   3.28   \\ \hline
50        &  0.60     &&    1.35  &&   3.76   \\ \hline
60        &  0.73     &&    1.64  &&   4.56   \\ \hline
70        &  2.02     &&    4.56  &&   12.7   \\ \hline
80        &  9.76     &&    22.0  &&   61.0  \\ \hline
90        &  11.4     &&    25.7  &&   71.3  \\ \hline \hline
\end{tabular}
\caption{Integrated luminosities for the combined $ll$ and $lj$ analysis needed to reach a 95 \% CL exclusion, 3$\sigma$ and 5$\sigma$ discovery for a Higgs boson with SM-like couplings to the fermions, at the LHC. }
\end{center}
\label{tab:lumi}
\vskip -0.8cm
\end{table}

The SM signal and all background processes were generated with CalcHEP~\cite{Pukhov:2004ca}. In the models with an extended scalar sector, signal cross sections were evaluated with FeynArts/FormCalc/ LoopTools~\cite{looptools}. As each $\tau$ can either decay leptonically or hadronically and a three jet final state is very hard to identify at a hadron collider, we will concentrate on the other two possibilities - two taus decaying leptonically ($ll$) or one tau decaying leptonically  and the other hadronically ($lj$).  In both analyses we have considered the main source of irreducible background, $pp \to Z/\gamma^* j$,  and the most relevant sources of reducible background, $pp\to W^+ W^- j$,  $pp\to W j j$, $pp \to t\bar{t}$ and the QCD background $jjj$. A very detailed discussion of the analysis can be found in~\cite{us}. In tab.~\ref{tab:lumi} we present the luminosities required for a 95 \% Confidence Level (CL) exclusion, 3$\sigma$ and 5$\sigma$ discovery of a Higgs boson with SM-like Higgs couplings to the fermions at $\sqrt{s} = 14$  TeV as a function of the Higgs mass. A light Higgs boson with SM-like couplings to the fermions can be excluded at 95 \% CL in the mass range 20--60 $GeV$ with less than 1 $fb^{-1}$ of total integrated luminosity.

\begin{figure}
\begin{center}
\includegraphics[width=6.95cm]{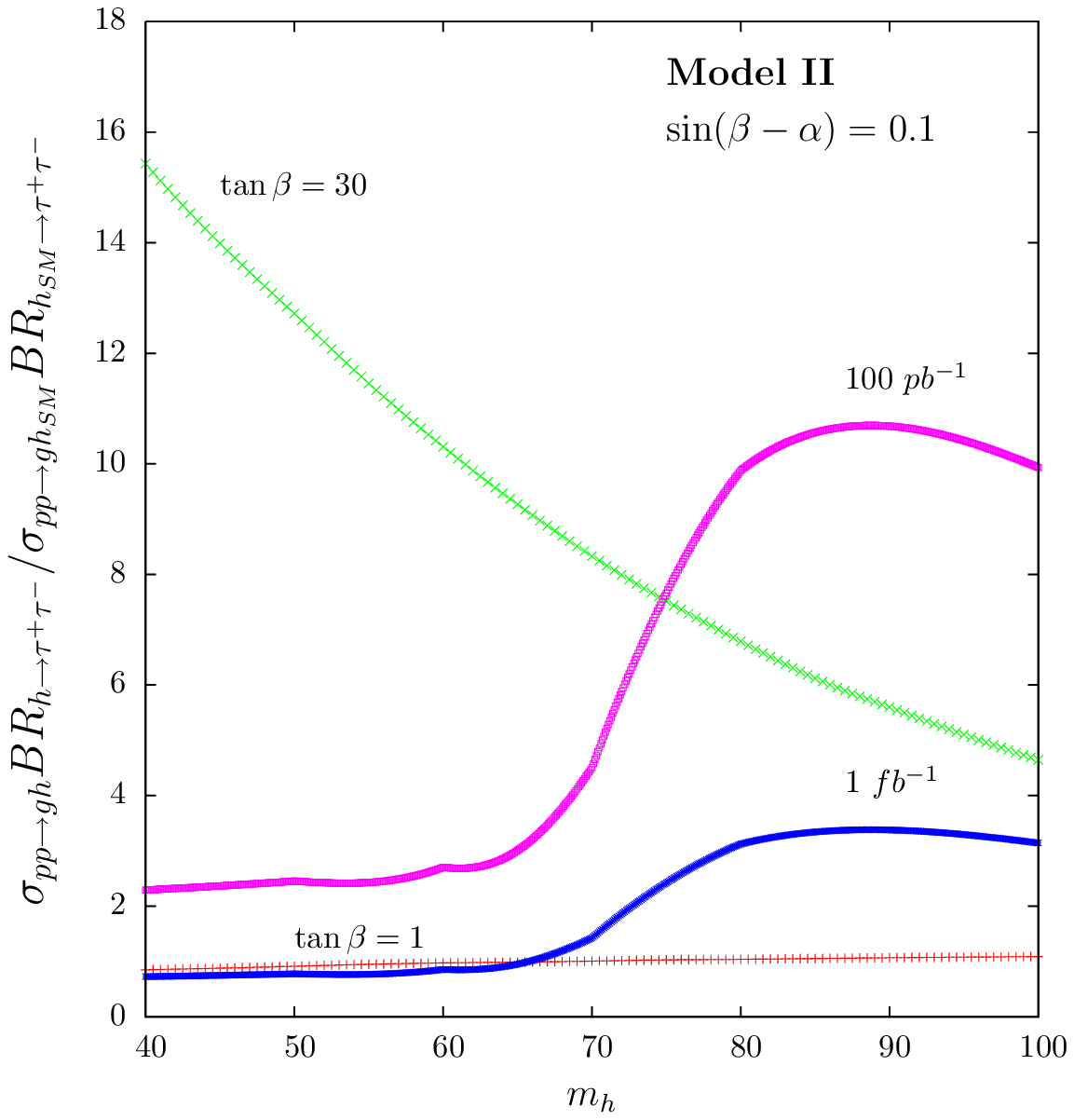}
\includegraphics[width=7.05cm]{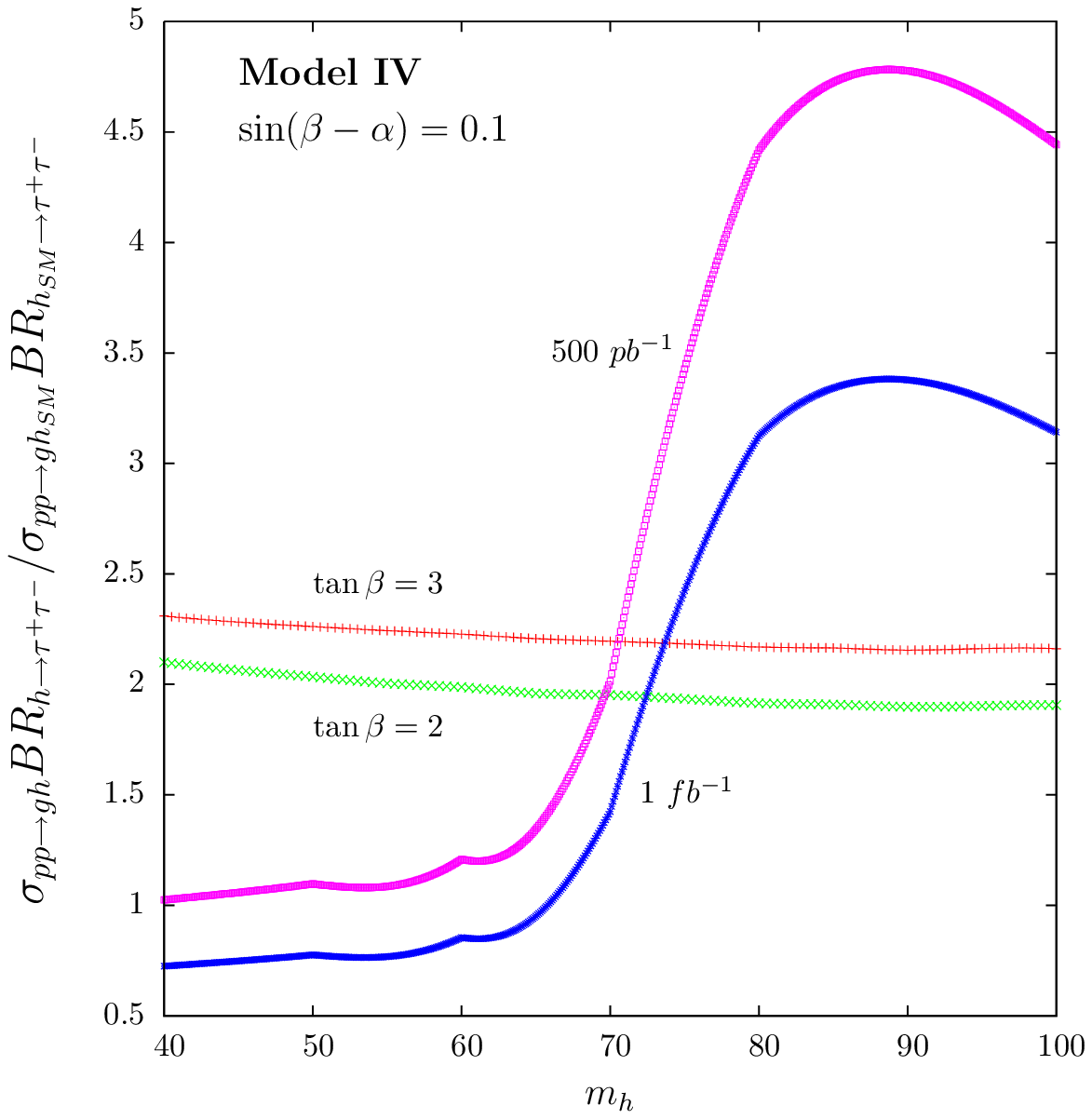}
\end{center}
\vspace{-0.8cm}
\caption{Left panel - ratio between $\sigma({pp \to hg}) \, {\rm{BR}}({h \to \tau^+ \tau^-})$ in model II and the SM $\sigma({pp \to h_{SM} g}) \, {\rm{BR}}({h_{SM} \to \tau^+ \tau^-})$ as a function of $m_h$ and $\tan \beta=$ 1 and 30. Right panel - same ratio now for model IV and $\tan \beta=$ 2 and 3. In both cases we take $\sin (\beta - \alpha)= 0.1$. We also show lines of total integrated luminosity $100 \, pb^{-1}$ and $1 \, fb^{-1}$ for model II and $500 \, pb^{-1}$ and $1 \, fb^{-1}$ for model IV.}
\label{fig:2HDM}
\vskip -0.3cm
\end{figure}

In fig.~\ref{fig:2HDM} we present the best scenarios for the pure 2HDM cases. The integrated luminosity lines represent the total luminosity needed to exclude the model at 95 \% CL. It is clear that there are regions of the parameter space that can be probed with less than $100 \, pb^{-1}$ of integrated luminosity with the LHC working at an energy of 14 TeV. The regions easily probed are for very light Higgs (mass below 60 GeV) and large $\tan \beta$ values in Model II and small to moderate values of $\tan \beta$ for Model IV.

\begin{figure}
\begin{center}
\includegraphics[width=6.9cm]{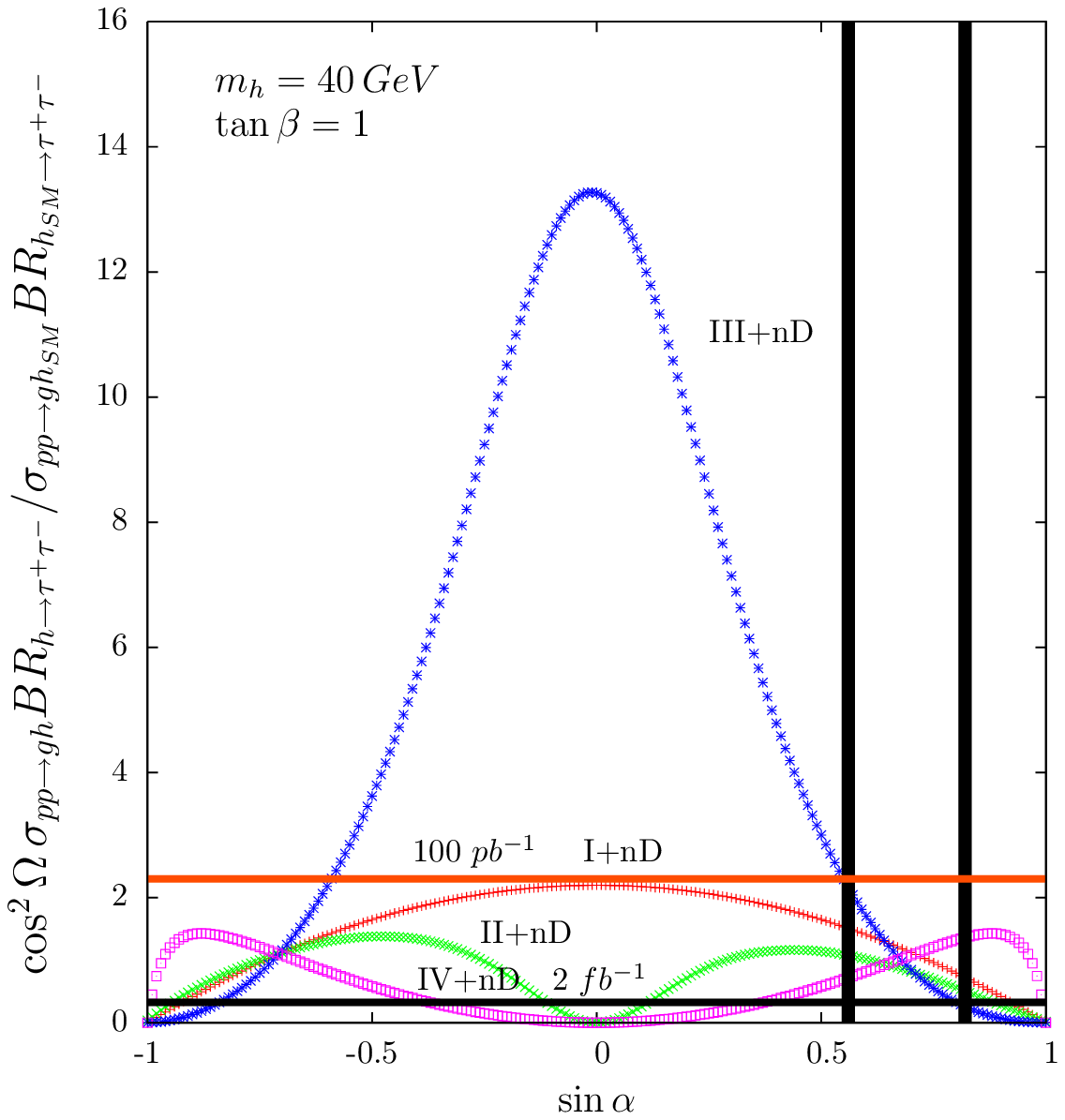}
\includegraphics[width=6.9cm]{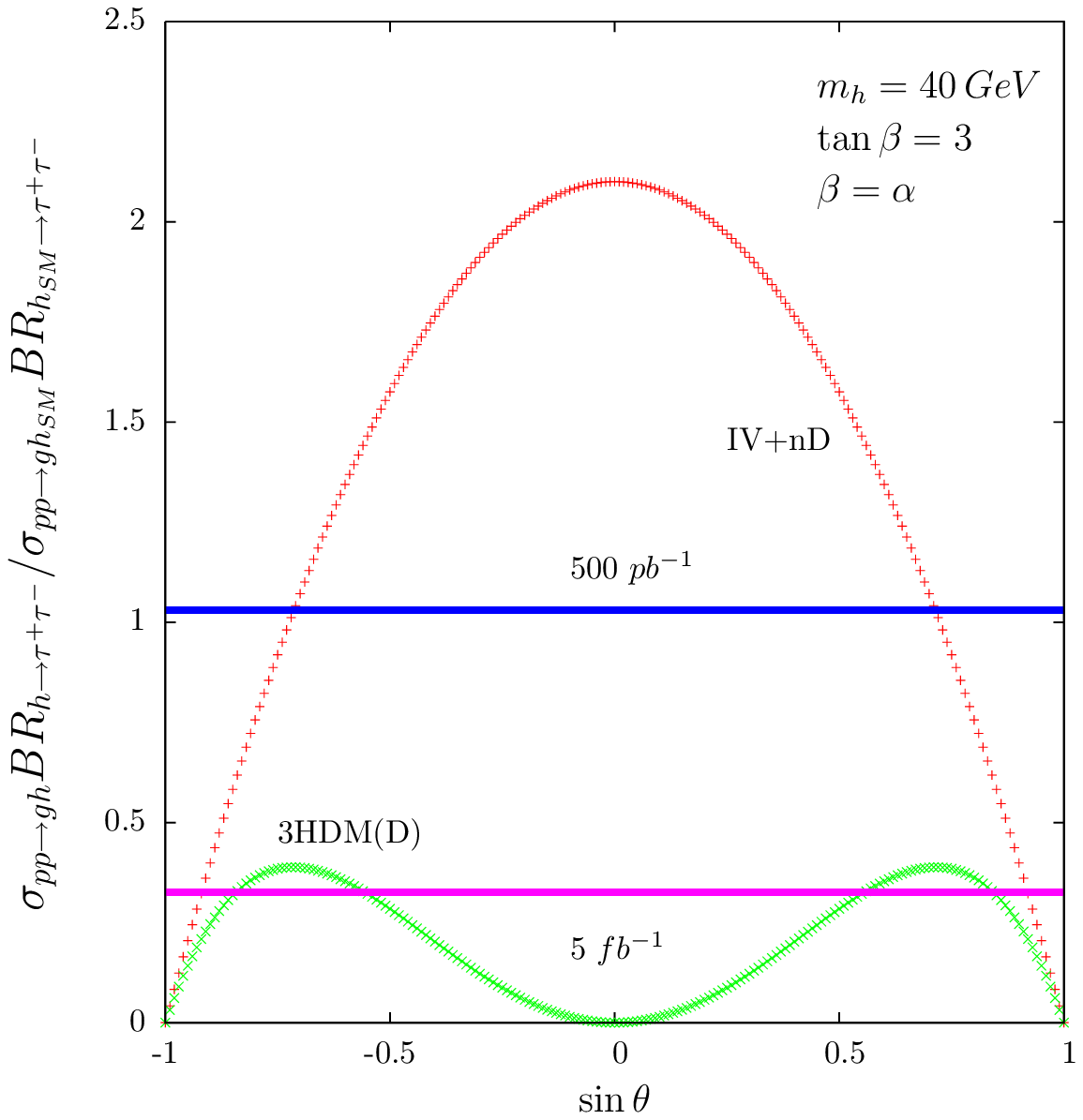}
\end{center}
\vspace{-0.8cm}
\caption{Left panel - ratio between $\sigma({pp \to hg}) \, {\rm{BR}}({h \to \tau^+ \tau^-})$ and the SM $\sigma({pp \to h_{SM} g}) \, {\rm{BR}}({h_{SM} \to \tau^+ \tau^-})$ multiplied by the factor $cos^2 \Omega$ for $m_h = 40 \, GeV$ and $\tan \beta=1$ for all 2HDM+nD. We also present the total integrated luminosities $100 \, pb^{-1}$ and $2 \, fb^{-1}$ Right panel - same ratio (without $cos^2 \Omega$) for 2HDMIV+nD and 3HDM(d) as a function of $\sin \theta$ and $\tan \beta= 3$ and $m_h = 40 \, GeV$.}
\label{fig:2HDMext}
\vskip -0.3cm
\end{figure}

More general models have first to respect the LEP bounds, that is, to have small couplings between the lightest Higgs and gauge bosons.  If almost no mixing occurs between the new doublet and the remaining ones and the new VEV is maximal ($\cos \Omega \ll 1$ and $\sin \theta \ll 1$)  all cross sections are rescaled, and therefore enhanced, as $\sigma^{\rm 2HDM} \to 1/\cos^2\Omega \, \sigma^{\rm 2HDM}$. Both $\alpha$ and $\tan \beta$ are now free to vary in all the allowed range,  provided theoretical and experimental constraints are fulfilled. In the left panel of fig.~\ref{fig:2HDMext} we present the ratio between $\sigma({pp \to hg}) \, {\rm{BR}}({h \to \tau^+ \tau^-})$ and the its SM equivalent, multiplied by $\cos^2 \Omega$ for all Yukawa extensions of the 2HDM+nD. Note that to obtain the actual value of the cross section one needs to multiply it by $1/\cos^2 \Omega$ and therefore the numbers will always be larger than the ones shown in the figures. Again, it is clear that $100 \, pb^{-1}$ are enough to constraint a big portion of model III+nD while with $2 \, fb^{-1}$ just marginal regions of the models are left untested. Between the two vertical lines, the allowed region of the parameter space for the pure 2HDM is shown. In the right panel, the scenario $\sin \Omega \ll 1$ and $\alpha \approx \beta$ for model IV+nD and 3HDM(D) is shown. By taking $\sin \Omega = 0.1$ and $\alpha = \beta$ we conclude that the prospect of excluding a light Higgs in model IV+nD is good but even with $5 fb^{-1}$ of integrated luminosity only a small portion of the 3HDM(D) will be probed at 95 \% CL.

A very interesting class of models is obtained by setting $g_{VVh} = 0$ which means $\sin (\beta - \alpha) =  -\tan \Omega \tan \theta$ and therefore it does  \textit{not require} any special limit to avoid the LEP bound. As an example, if $\sin (\beta - \alpha) =  \tan \Omega = - \tan \theta =  1$ the LEP bound is avoided and the lightest Higgs from 3HDM(D) will have SM-like couplings to fermions. Finally, other scenarios where the LEP bounds are avoided but add nothing relevant to the previous discussion are: $\sin \theta << 1$  and $\alpha \approx \beta$; $\sin \Omega << 1$ and $\cos \theta << 1$.

With the LHC running and with the search for the Higgs boson on the way, we should ask ourselves what to do if we do not find a SM Higgs boson. It seems clear that we should turn our attention to more general potentials and in particular to the ones where a light Higgs boson is allowed. However, even if a Higgs boson is found and even if it looks very much like the SM Higgs boson, we should make sure that we did not miss any other (pseudo)scalar particle potentially present in the data. We believe this work is a very important contribution to achieve such a goal.


\begin{thebibliography}{99}

\bibitem{Schael:2006cr}
  S.~Schael {\it et al.}  [ALEPH, DELPHI, L3 and OPAL Collaborations],
  Eur.\ Phys.\ J.\  C {\bf 47} (2006) 547.



\bibitem{Ellis:1987xu}
  R.~K.~Ellis, I.~Hinchliffe, M.~Soldate and J.~J.~van der Bij,
  Nucl.\ Phys.\  B {\bf 297} (1988) 221.



\bibitem{catalogue}
V.~D.~Barger, J.~L.~Hewett and R.~J.~N.~Phillips,
  Phys. \ Rev.\  D {\bf 41} (1990) 3421.


\bibitem{Barger:2009me}
  V.~Barger, H.~E.~Logan and G.~Shaughnessy,
  Phys.\ Rev.\  D {\bf 79} (2009) 115018.

\bibitem{Grossman:1994jb}
  Y.~Grossman,
  Nucl.\ Phys.\  B {\bf 426} (1994) 355.
\bibitem{Pukhov:2004ca}
  A.~Pukhov,
  arXiv:hep-ph/0412191.


\bibitem{looptools}
T.~Hahn, Comput.\ Phys.\ Commun.\  {\bf 140} (2001)  418;
T.~Hahn, C.~Schappacher,
Comput.\ Phys.\ Commun.\  {\bf 143} (2002)  54;
T.~Hahn, M.~Perez-Victoria,
Comput.\ Phys.\ Commun.\  {\bf 118} (1999)  153;
G.~J.~van Oldenborgh,
Comput.\ Phys.\ Commun.\  {\bf 66} (1991) 1;
T.~Hahn, Acta Phys.\ Polon.\ B {\bf 30} (1999)  3469.

\bibitem{us}
  A.~Belyaev, R.~Guedes, S.~Moretti and R.~Santos,
  arXiv:0912.2620 [hep-ph];
  A.~Belyaev, R.~Guedes, S.~Moretti and R.~Santos,
  Phys.\ Rev.\  D {\bf 81} (2010) 095006.

\end{thebibliography}
\end{document}